# An asymptotic formula for displacement field in triangular lattice with vacancy

V.A. Tsaplin[1], V.A. Kuzkin[1,2]

vtsaplin@yandex.ru, kuzkinva@gmail.com

[1]Institute for Problems in Mechanical Engineering, Bolshoy pr. V.O. 61, St. Petersburg, 199178, Russia

[2]Peter the Great Saint Petersburg Polytechnic University, Polytechnicheskaya st. 29, St. Petersburg, 195251, Russia

**Abstract**

A harmonic triangular lattice with a vacancy under imposed volumetric strain is considered. Simple asymptotic formula for the displacement field is derived. The formula has reasonable accuracy at all lattice nodes. Strain concentration factor, defined as the ratio of the maximal deformation of the bonds adjacent to the vacancy to the deformations of bonds at infinity, is calculated. It is shown that the asymptotic formula predicts the strain concentration factor with 4% accuracy. The results are compared with predictions of continuum elasticity theory. Effective diameter of the vacancy is calculated.

## 1. Introduction

Recent advances in nanotechnologies allow to create crystals with very low defect concentrations and unique mechanical properties [1-8]. Prediction of these properties is one of the key problems for modern mechanics of materials. This problem has been studied in literature from both continuum and discrete points of view. Continuum theory of lattice defects have been developed in pioneering works of Eshelby [9]. In continuum mechanics, defects are modelled as pores in a homogeneous elastic medium, and continuum mechanics tools are used for calculation of displacement fields, elastic interaction of defects, etc. Continuum mechanics modelling is expected to be appropriate for calculation of displacement fields far from the defects. Near defects where the discreteness plays important role, continuum modelling may become inadequate. For example, in paper [10] it is shown that the effect of discreteness is significant even in the absence of surface tension.

Exact displacement fields around defects can be calculated using harmonic crystal model. This model is widely used for description of elastic properties [11], interaction of defects [9], crack propagation [12], heat transfer [13], transient thermal processes [14] and other phenomena. Several discrete methods for calculation of displacement fields in crystals have been proposed in literature [15-17]. Two statements of the problem are considered. In the framework of the lattice statics [15] and the lattice Green's function methods [16], the displacement field is generated by additional forces acting on atoms. The forces model nonlocality of interatomic interactions in a crystal. Alternatively, the displacement field can be represented via the mean strain of the periodic cell containing a vacancy (see e.g. paper [17]). The. In both cases, an exact solution is obtained using the discrete Fourier transform.

The discrete Fourier transform yields the displacement field in a form of sums (for periodic systems) or integrals (for infinite systems). These integrals are usually quite complicated [17]. Therefore analysis of the exact solutions is not straightforward. However the integrals contain large parameter, notably the particle index proportional to distance from the vacancy. Then expansion with respect to the large parameter can be carried out using asymptotic methods [18,19]. The asymptotic expansion yields simple closed-form expressions converging to the exact solution with increasing the large parameter. As it is shown below, the convergence can be quite fast.

In the present paper, we consider asymptotic behavior of an exact displacement field obtained in paper [17] for triangular lattice with vacancy under imposed volumetric strain. Simple asymptotic expression for the displacement field is derived. The expression allows to make a link between discrete and continuum descriptions of elastic fields caused by a vacancy. In particular, effective diameter of the vacancy and strain concentration are estimated.

## 2. Displacement field in triangular lattice with vacancy: an exact solution

Consider an infinite triangular lattice with a single vacancy (see figure 1). The lattice consists of identical particles. The nearest neighbors are connected by linear elastic springs. Infinitesimal deformations of the lattice are considered. Displacements of the particles under imposed volumetric strain $\varepsilon$ are calculated.

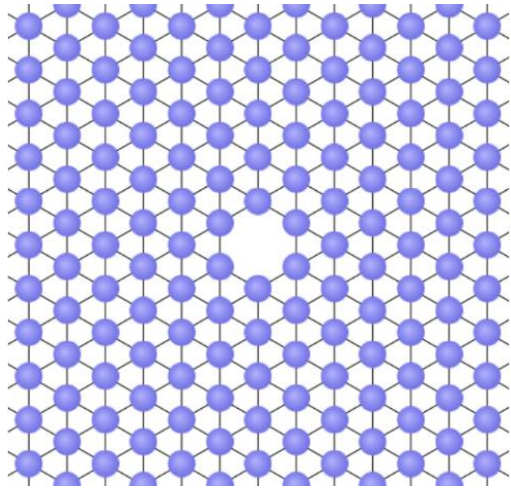

Figure 1. Triangular lattice with a vacancy.

We introduce two indices $n, k$ that enumerate the particles so that their position vectors are:

$$\boldsymbol{r}_{n,k} = a(n\boldsymbol{e}_1 + k\boldsymbol{e}_2), \quad \boldsymbol{e}_1 = \frac{\sqrt{3}}{2}\boldsymbol{i} + \frac{1}{2}\boldsymbol{j}, \quad \boldsymbol{e}_2 = -\frac{\sqrt{3}}{2}\boldsymbol{i} + \frac{1}{2}\boldsymbol{j}, \quad (1)$$

where $\boldsymbol{e}_1$, $\boldsymbol{e}_2$ are unit vectors that correspond to directions of bonds in the lattice; $\boldsymbol{i}$, $\boldsymbol{j}$ are unit vectors of Cartesian basis (in fig. 1: $\boldsymbol{i}$ is horizontal); $a$ is the equilibrium distance between the nearest neighbors.

Particle displacements are represented as

$$\boldsymbol{u}_{n,k} = \frac{2}{3}\Big((2u_{n,k} + v_{n,k})\boldsymbol{e}_1 + (2v_{n,k} + u_{n,k})\boldsymbol{e}_2\Big), \quad (2)$$
$$u_{n,k} = \boldsymbol{u}_{n,k} \cdot \boldsymbol{e}_1, \quad v_{n,k} = \boldsymbol{u}_{n,k} \cdot \boldsymbol{e}_2.$$

Equilibrium of the lattice is described by difference equations with respect to particle displacements. An exact analytical solution of these equations is obtained in paper [17]. In the case of volumetric deformation, the exact solution reads

$$u_{n,k} = v_{k,n} = a\varepsilon\left(n - \frac{k}{2} - \frac{G_{n,k}}{8\pi^2 + G_{1,0}}\right),$$

$$G_{n,k} = \int_{-\pi}^{\pi}\int_{-\pi}^{\pi} g(x,y) \sin(nx + ky)\, dx\, dy, \quad (3)$$

$$g(x,y) = \frac{\sin y \sin^2\frac{x}{2} - \sin x \sin^2\frac{x+y}{2}}{\sin^2\frac{x}{2}\sin^2\frac{y}{2} + \sin^2\frac{x}{2}\sin^2\frac{x+y}{2} + \sin^2\frac{y}{2}\sin^2\frac{x+y}{2}},$$

where $\varepsilon$ is the volumetric mean strain.

Formula (3) shows that displacements depend on integral $G_{n,k}$. The integrand of $G_{n,k}$ changes sign and oscillates with frequency proportional to particle indexes $n, k$. Then asymptotic expansion for $|n|, |k| \to +\infty$ can be carried out. From physical point of view, the expansion yields the behavior of the displacement field far from the vacancy ($r_{n,k} \to \infty$).

## 3. An asymptotic formula for the displacement field

We investigate asymptotic behavior of the displacement field (integral $G_{n,k}$) far from the vacancy.

Consider the case $k = 0$. Then the integral $G_{n,o}$ reads

$$G_{n,0} = \int_{-\pi}^{\pi} f(x) \sin nx\, dx, \quad f(x) = \int_{-\pi}^{\pi} g(x,y)\, dy \quad (4)$$

We calculate asymptotics of integral (4) for large $n$. Function $f(x)$ is $2\pi$-periodic and continuous with all derivatives on the whole period, except for the point $x = 0$. At $x = 0$ the function has a discontinuity (jump). We show that the leading term in asymptotic expansion of the integral $G_{n,0}$ for large $n$ is determined by the magnitude of this jump. Double integration by parts in formula (4) yields

$$G_{n,0} = -\frac{1}{n} f(x)\big|_{+0}^{2\pi-0} - \frac{1}{n^2}\int_0^{2\pi} f''(x) \sin nx\, dx \approx$$

$$\approx \frac{1}{n}\lim_{\varepsilon \to 0}\big(f(\varepsilon) - f(-\varepsilon)\big). \quad (5)$$

Here the integral term is neglected for the following reasons. The integrand is bounded and therefore the integral decays as $1/n^2$ or faster.

We calculate the jump of function $f$ at the point $x = 0$ using the identity

$$f(x) = \int_{-\pi}^{-\varepsilon} g(x,y)dy + \int_{-\varepsilon}^{\varepsilon} g(x,y)dy + \int_{\varepsilon}^{\pi} g(x,y)dy, \qquad (6)$$

where $\varepsilon \ll 1$. The function $g(x,y)$, defined by the expression (3), is regular in the integration domain except for the point $x = 0$, $y = 0$. Therefore the jump of function $f(x)$ at $x = 0$ is equal to the jump of the second term in formula (6). In the vicinity of the point $x = 0$, $y = 0$, function $g(x,y)$ can be approximated by the following formula:

$$g(x,y) \approx \frac{-4x}{x^2 + y^2 + xy}. \qquad (7)$$

This formula is obtained by the power series expansion of the numerator and denominator of function $g(x,y)$.

The jump of function $f$ at $x = 0$ is calculated by substitution of formula (7) into (6). Then formula (5) yields asymptotics of the integral $G_{n,0}$ for large $n$:

$$G_{n,0} \approx -\frac{16\pi}{\sqrt{3}n}. \qquad (8)$$

Using formula (8) we estimate the value $G_{1,0} \approx -29.02$. Note that the value calculated using exact formula (3) is $G_{1,0} \approx -30.87$, which is only 6% larger.

Consider the case $k \neq 0$. We introduce the new variable $z = y + nx/k$. Then integral $G_{n,k}$ in formula (3) takes the form:

$$G_{n,k} = \int_{-\pi}^{\pi} \tilde{f}(z) \sin kz \, dz,$$

$$\tilde{f}(z) = \int_{-\pi}^{\pi} \tilde{g}(x,z)dx, \quad \tilde{g}(x,z) = g(x,y). \qquad (9)$$

Then the problem is equivalent to the case $k = 0$ considered above (see equation (4)). It is assumed that $s = n/k$ is constant, therefore the condition $|k| \to \infty$ is equivalent to $n^2 + k^2 \to \infty$. Calculating the asymptote of the integral $G_{n,k}$, as described above, yields

$$n^2 + k^2 \to \infty: \quad G_{n,k} \approx \frac{8\pi}{\sqrt{3}} \frac{k - 2n}{n^2 - nk + k^2}. \qquad (10)$$

Formula (10) coincides with previous result (8) for $k = 0$. Also from formula (11) it follows that $G_{n,2n} = 0$. The same is true for the exact solution (3).

Substitution of formula (10) into (3) yields the asymptotic expression for displacement field in triangular lattice with vacancy subjected to the volumetric strain

$$\boldsymbol{u}_{n,k} \approx \varepsilon\left(1 + \frac{2}{\sqrt{3}\pi - 2} \frac{a^2}{r_{n,k}^2}\right) \boldsymbol{r}_{n,k}, \qquad (11)$$

where $\boldsymbol{r}_{n,k}$ is defined by formula (1); $r_{n,k} = |\boldsymbol{r}_{n,k}|$. Formula (11) shows, in particular, that the contribution of vacancy to the displacement field decays inversely proportional to distance from the vacancy.

## 4. Comparison with exact displacement field and predictions of continuum theory

We check the accuracy of formula (11) by comparison with exact displacement field (3). Radial components of displacements calculated using asymptotic formula (11) and exact formula (3) are shown in figure 2. In both cases, the displacement field $\tilde{\boldsymbol{u}}_{n,k} = \varepsilon \boldsymbol{r}_{n,k}$ corresponding to uniform deformation of the lattice is subtracted.

The main difference between asymptotic and exact displacement fields is that in the latter case the dependence $u_r(r/a) = \boldsymbol{u}_{n,k} \cdot \boldsymbol{r}_{n,k}/r_{n,k}$ is nonmonotonic. The maximum relative error of asymptotic formula (11) for $r/a = 2$ is 15%. For all other points, the error is less than 12%.

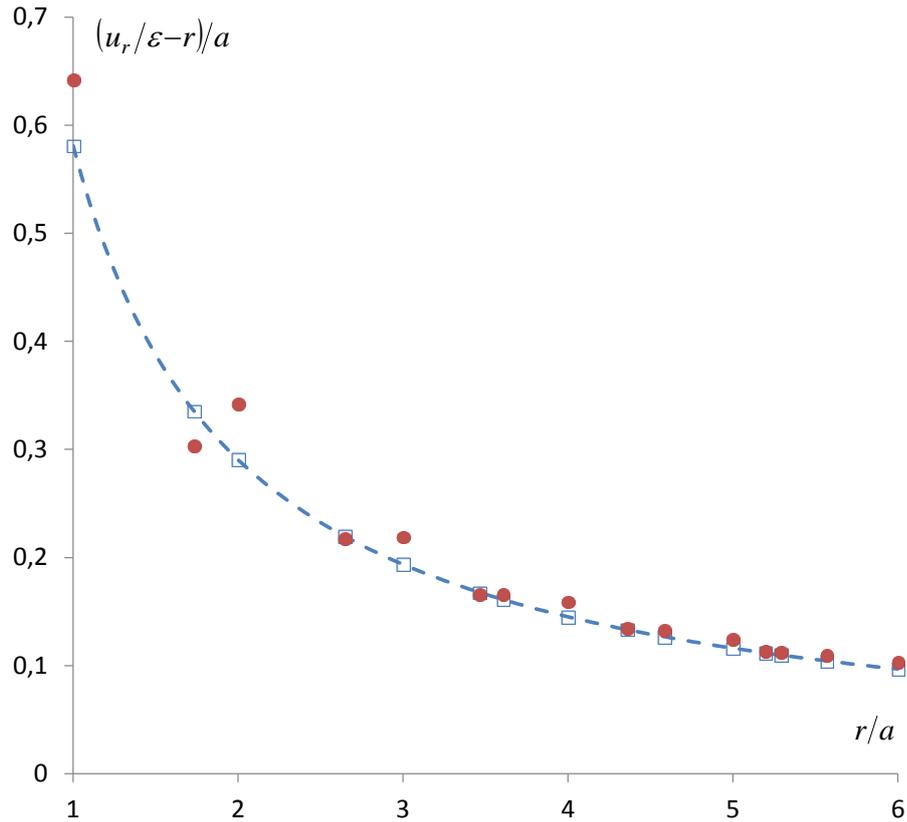

Figure 2. Displacements caused by a vacancy in a triangular lattice under remotely applied volumetric strain $\varepsilon$. Squares — asymptotic formula (11), circles — exact solution (3).

Compare the results with continuum linear elasticity theory. Assume that the vacancy can be modeled by a circular hole in an infinite plate. Displacements field in a plate with the Poisson ratio equal to 1/3 containing a circular hole of diameter $d$ loaded by remotely applied hydrostatic stress $\sigma_0$ (Kirsch problem) reads

$$\boldsymbol{u}(\boldsymbol{r}) = \varepsilon\left(1 + \frac{d^2}{2r^2}\right)\boldsymbol{r}, \qquad \varepsilon = \frac{\sigma_0}{2K}, \qquad (12)$$

where $K$ is the bulk modulus. It is seen that the continuum displacement field (12) has the same form as asymptotic formula (11) for the discrete displacement field.

Comparison of formulas (11), (12) yields the following expression for diameter of the hole modeling vacancy

$$\frac{d}{a} = \frac{2}{\sqrt{\sqrt{3}\pi - 2}} \approx 1.08. \tag{13}$$

This value is in a good agreement with the result obtained in paper [17] by numerical fitting of the exact displacement field (3).

Formula (13) has simple geometrical interpretation. The triangular lattice is usually visualized by close-packing of equal spheres. Diameter of a sphere is equal to equilibrium distance $a$. Formula (13) shows that effective diameter of the hole modelling a vacancy is only 8% larger than diameter of the sphere.

In continuum theory, the presence of the hole leads to concentration of stresses. The stress concentration factor for a circular hole under imposed volumetric strain is equal to 2 at all points of the boundary; for other shapes, the maximal, around the boundary, concentration factor is higher. Vacancy causes similar effect; however, calculation of stress concentration factor is not straightforward, since stresses in discrete media are defined ambiguously [20]. Therefore we consider the strain concentration factor $k$ defined as the ratio of the maximal deformation of the bonds adjacent to the vacancy to the deformations of bonds at infinity. In paper [17], the value $k \approx 1.64$ was obtained using an exact displacement field (3). The asymptotic formula (11) yields $k \approx 1.58$, which is only 4% smaller. Note that both values are substantially lower than prediction of the continuum elasticity theory.

Thus asymptotic formula (11) can be used as approximation of the displacement field at all points of the lattice.

## 5. Conclusions

Simple asymptotic expression (11) for the displacement field in harmonic triangular lattice under imposed volumetric strain was derived. The expression has reasonable accuracy for all lattice nodes and it converges to the exact solution with increasing distance from the vacancy. In particular, the expression allows to calculate the strain concentration factor with 4% accuracy.

Comparison with predictions of continuum elasticity (Kirsch problem) was carried out. It was shown that the asymptotic expression has the same form as displacement field in a stretched continuum plate with circular hole. Effective diameter of the vacancy was estimated. Therefore continuum elasticity theory correctly predicts asymptotic behavior of the displacement field caused by the vacancy. However the influence of vacancies on strength of the crystal can not be accurately modeled by continuum theory. For further discussion, we refer to paper [17].

**Acknowledgements**

The authors are grateful to A.M. Krivtsov and M.L. Kachanov for useful discussions.